\documentclass[12pt]{article}
\usepackage[italian,english]{babel}
\usepackage{graphicx}
\usepackage{psfig,picinpar,floatflt,epsf}
\usepackage{latexsym}
\usepackage{amsmath,amsfonts,amssymb,amsthm,wasysym}
\def\bseq{\begin{subequation}}  
\def\eseq{\end{subequation}}
\def\bsea{\begin{subeqnarray}}  
\def\esea{\end{subeqnarray}}

\newcommand{\bbox}{\lower.2ex\hbox{$\Box$}}

\newcommand{\beq}{\begin{equation}}
\newcommand{\eeq}{\end{equation}}
\newcommand{\bea}{\begin{eqnarray}}
\newcommand{\eea}{\end{eqnarray}}
\newcommand{\ena}{\end{eqnarray}}

\newcommand {\non}{\nonumber}

\newcommand{\f}{\phi}

\newcommand{\p}{\pi}

\newcommand{\Tr}{{\rm Tr}}

\renewcommand{\[}{\left[}

\newcommand{\be}{\begin{equation}}
\newcommand{\ee}{\end{equation}}
\newcommand{\mc}{\mathcal}
\def\tr{{\rm tr}\ }

\begin{document}
\begin{titlepage}
{\hbox to\hsize{August 2006 \hfill}}
\begin{center}
\vglue .06in
\vskip 40pt

{\Large\bf Non-supersymmetric Meta-Stable vacua} \\ 
[.07in]
{\Large\bf in $SU(N)$ SQCD with adjoint matter} 
\\[.7in]

{\large\bf A.Amariti\footnote{antonio.amariti@mib.infn.it},
L. Girardello\footnote{luciano.girardello@mib.infn.it} and
A. Mariotti\footnote{alberto.mariotti@mib.infn.it}
}
\\[.5in]
{\it Dipartimento di Fisica, Universit\`a degli Studi di
Milano-Bicocca\\ 
and INFN, Sezione di Milano-Bicocca, piazza delle Scienze 3, I 20126 Milano, Italy}
\\[.8in]

{\bf ABSTRACT}\\[.3in]
\end{center}
We investigate models of $SU(N)$ SQCD with adjoint matter
and non trivial mesonic deformations. 
We apply standard methods in the dual magnetic theory and we find 
meta-stable supersymmetry breaking vacua with
arbitrary large lifetime. We comment on the difference
with known models.

\vskip 30pt
${~~~}$ \newline

\end{titlepage}

\section{Introduction}
Long living meta-stable vacua breaking supersymmetry exist in classes of $\mc{N}=1$
gauge theories 
of the SQCD type with massive fundamental
matter \cite{Rattazzi,Seiberg}.

The novelty of the approach of \cite{Seiberg} relies on theories for which Seiberg-like
duality exists i.e. (electric) theories which are asymptotically free in the ultraviolet and
strongly coupled in the infrared, where the physics can be described
in terms of weakly coupled dual (magnetic) theories (for reviews see \cite{intriligator,Strassler}).
In the region of small fields this dual description
can be studied as a model of
pure chiral fields.
Supersymmetry is broken by the  
rank condition \cite{Seiberg}, i.e. not all
the $F$-term conditions can be satisfied.
Roughly speaking the next 
step is to recover, in this magnetic infrared, a generalized chiral
O'Raifeartaigh model with supersymmetry breaking vacua.

These non supersymmetric vacua have typically classical flat
directions which can be lifted by quantum corrections. In \cite{Seiberg} it
has been proved that such corrections generate positive mass terms for the pseudomoduli
leading to long lived metastable vacua.
These facts
should be tested in different supersymmetric theories.
Some generalization have already appeared upholding the notion that such phenomenon
is rather generic \cite{Uranga,ooguri,Ray}. 
The relative stability of the vacua is a rather delicate
issue. Remarks about the corresponding string configurations
corroborating the stability analysis have also appeared \cite{ooguri2,Uranga2,Uranga3,Braun}.

In this paper we study theories with 
adjoint chiral fields with cubic superpotential \`a la KSS 
\cite{Kutasov1,Kutasov2,Kutasov3}.
Such superpotentials generate a further meson in the dual magnetic theory:
this might produce several pseudogoldstone excitations
and jeopardize the 1-loop stability of the
non supersymmetric vacua. There must be 
enough $F$ and/or $D$ equations to
give tree level masses.
A viable model, of string origin, with 
two gauge groups has been presented in \cite{ooguri}.

We consider a theory with one gauge group $SU(N_c)$ and two massive 
electric adjoint fields, where the most massive one gets integrated out.
This amounts to add a massive 
mesonic deformation in the dual theory. This avoids dangerous extra flat directions
which cannot be stabilized at 1-loop.
A discussion of the possible interpretation
via D-brane configurations can be found in \cite{Giveon,Giveon1}.

In the study of the magnetic dual theory we find a tree-level non supersymmetric vacuum
which is stabilized by quantum corrections; we show that this is a metastable state 
that decays to a supersymmetric one after a parametrically long time.  
A landscape of non supersymmetric metastable vacua, present at classical level,
 disappears at quantum level. 
Differently from \cite{Seiberg,Uranga,ooguri} in our model there is no
$U(1)_R$ symmetry 
and our minimum will not be at the origin of the field space,
making our computation much involved. We 
present most of our
results graphically, giving analytic expressions in some sensible limits.
We follow the computational strategy of \cite{Seiberg}.

In section 2 we recall some basic elements of the KSS duality and introduce 
the model that we consider through the paper.
In section 3 we solve the the $D$ and $F$ equations finding
an energy local minimum
where supersymmetry
is broken by a rank condition.
In section 4 we compute the 1-loop effective potential around this 
vacuum 
and find that it is stabilized by the quantum corrections.
In section 5 we restore supersymmetry by non perturbative gauge dynamics
and recover supersymmetric vacua. Using this result we estimate
the lifetime of our metastable vacuum in section 6.

\section{$\mc{N}=1$ SQCD with adjoint matter}
Here we introduce some useful elements about electric/magnetic duality for
supersymmetric gauge theory with an adjoint field \cite{Kutasov1,Kutasov2,Kutasov3}.
We consider $\mc{N}=1$ supersymmetric $SU(N_c)$ Yang Mills theory coupled
to $N_f$ massive flavours ($Q^{i}_{\alpha},\tilde Q^{ j \beta}$) 
in the fundamental and
antifundamental representations of the gauge group ($\alpha,\beta=1,\dots N_c$) and 
in the antifundamental and
fundamental representations of the flavour group ($i,j=1,\dots N_f$), respectively.
We also consider a charged chiral massive adjoint superfield $X_{\beta}^{\alpha}$ 
with superpotential%
\footnote{($\Tr$) means tracing on the color indices, while ($\tr)$ on the flavour ones.}
\be
\label{supel}
W_{el}=\frac{g_X}{3} \Tr X^3+\frac{m_{X}}{2} \Tr X^2 +\lambda_X \Tr X 
\ee
where $\lambda_X$ is a Lagrange multiplier 
enforcing
the tracelessness condition $\Tr X=0$.
The Kahler potential for all the fields is taken to be
canonical.
This theory is asymptotically free in the range $N_f< 2 N_c$ and it admits stable vacua
for $N_f>\frac{N_c}{2}$ \cite{Kutasov2}.

The dual theory \cite{Kutasov1,Kutasov2,Kutasov3} 
is $SU(2 N_f- N_c \equiv \tilde N)$ with $N_f$ magnetic flavours ($q,\tilde q$), a magnetic
adjoint field $Y$ and two gauge singlets build from electric mesons ($M_1= Q\tilde Q$, $M_2=Q X \tilde Q$),
with magnetic superpotential 
\be
\label{supma}
W_{magn}=\frac{\tilde g_Y}{3} \Tr Y^3+\frac{\tilde m_Y}{2} \Tr Y^2+\tilde \lambda_Y \Tr Y 
-\frac{1}{\mu^2} \tr \left(\frac{\tilde m_Y}{2} M_1 q \tilde q+\tilde g_Y M_2 q \tilde q 
+ \tilde g_Y M_1 q Y \tilde q \right) 
\ee
where the relations between the magnetic couplings and the electric ones are
\be
\label{couplings}
\tilde g_Y=-g_X , \qquad \tilde N \tilde m_Y=N_c m_X.
\ee
The intermediate scale $\mu$ takes into account
the mass dimension of the mesons in the dual description. 
The matching between the microscopic scale ($\Lambda$) and the macroscopic scale ($\tilde \Lambda$)
is
\be
\label{matching}
\Lambda^{2N_c-N_f} \tilde \Lambda^{2 \tilde N-N_f}=\left( \frac{\mu}{g_X} \right)^{2 N_f} \,.
\ee
We look for a magnetic infrared free regime in order to rely on 
perturbative computations at low energy. The $b$ coefficient of the beta function
is $b=(3 \tilde N-N_f)-\tilde N$, negative  for $N_f<\frac{2}{3} N_c$ and so we will 
consider the window for the number of flavours 
\be
\label{range}
\frac{N_c}{2}<N_f<\frac{2}{3}N_c \qquad \Rightarrow \qquad 0<2 \tilde N < N_f 
\ee
where the magnetic theory
is IR free and it admits stable vacua.

\subsection{Adding mesonic deformations}
We now add to the electric potential (\ref{supel}) the gauge singlet
deformations
\be
\label{deltaW}
W_{el} \to W_{el}+\Delta W_{el}\qquad \Delta W_{el}= \lambda_Q \, \tr Q X \tilde Q  +m_Q \, \tr Q \tilde Q
+ h \, \tr (Q \tilde Q)^2
\ee
The first two terms are standard deformations of the electric
superpotential that
don't spoil the duality 
relations (e.g. the scale matching condition (\ref{matching})) \cite{Kutasov3}.
The last term of (\ref{deltaW}) 
can be thought as originating from a second largely massive adjoint field $Z$ in the 
electric theory with superpotential
\be
W_{Z}=m_Z \Tr Z^2+ \Tr \, Z Q \tilde Q 
\ee
and which has been integrated 
out \cite{Giveon,Giveon1,cachazo}.
The mass $m_{Z}$ has to be considered larger than $\Lambda_{2A}$, the strong scale of the
electric theory with two adjoint fields.
This procedure leads to the scale matching relation
\be
\label{scaleadj}
\Lambda_{2A}^{N_c-N_f}=\Lambda_{1A}^{2 N_c-N_f} m_Z^{-N_c}
\ee
where $\Lambda_{2A}$ and $\Lambda_{1A}$ are the strong coupling scales 
before and after having integrated out the adjoint field $Z$, i.e.
with two or one adjoint fields respectively.

The other masses in this theory have to be considered 
much smaller than the strong scale $\Lambda_{2A}\gg m_Q,m_X$. 
This forces, via (\ref{scaleadj}), the scale $\Lambda_{1A}$ and the masses 
to satisfy the relations
\be
\label{rangemasses}
\frac{m_Q m_Z}{\Lambda_{1A}^2} \ll 1 \qquad \qquad  \frac{m_X m_Z}{\Lambda_{1A}^2} \ll 1
\ee
We will work in this range of parameters in the whole paper,
translating these inequalities in the dual (magnetic) context.

We also observe that in (\ref{scaleadj}) the coefficient $b$ of the beta function 
for the starting electric theory with two adjoint fields is
$b=N_c-N_f$ and the theory is asymptotically free
for $N_f<N_c$. This range is still consistent with our magnetic IR free
window (\ref{range}).
The dimensional coupling $h$ in our effective theory (\ref{deltaW})
results $h=\frac{1}{m_Z}$ so it must be thought as a small 
deformation.
In analogy with \cite{cachazo}\footnote{Where it was done in the context of Seiberg duality.} 
we can suppose that when $h$ is small
the duality relations are still valid and
obtain the full magnetic superpotential 
\bea
\label{supmagn}
W_{magn}&=&\frac{\tilde g_Y}{3} \Tr Y^3+\frac{\tilde m_Y}{2} \Tr Y^2+\tilde \lambda_Y \Tr Y 
-\frac{1}{\mu^2}\, \tr \left(\frac{\tilde m_Y}{2} M_1 q \tilde q+\tilde g_Y M_2 q \tilde q 
+ \tilde g_Y M_1 q Y \tilde q \right) \non \\
&&+ \lambda_Q~ \tr M_2  + m_Q~ \tr M_1 + h ~ \tr (M_1)^2 
\eea
For this dual theory the scale matching relation is the same as (\ref{matching})
with $\Lambda \equiv \Lambda_{1A}$ defined in (\ref{scaleadj}).\\
We consider the free magnetic range (\ref{range}), where the metric on 
the moduli space is smooth around the origin.
The Kahler potential is thus regular and has the canonical form 
\be
\label{kahler}
K=\frac{1}{\alpha_1^2 \Lambda^2} \tr M_1^{\dagger}  M_1+\frac{1}{\alpha_2^2 \Lambda^4} \tr M_2^{\dagger}  M_2+
\frac{1}{\beta^2} \Tr Y^{\dagger} Y+\frac{1}{\gamma^2} (\tr q^{\dagger} q+\tr \tilde q^{\dagger} \tilde q)
\ee
where ($\alpha_i,\beta,\gamma$) are unknown positive numerical coefficients.

\section{Non supersymmetric metastable vacua}
We solve 
the equations of motion for the chiral fields
of the macroscopic description
(\ref{supmagn}).
We will find a non supersymmetric vacuum in the region of small fields where the
$SU(\tilde N)$ gauge dynamics is decoupled.
The gauge dynamics becomes relevant in the large field region where it restores
supersymmetry via non perturbative effects (see sec.5).

We rescale the magnetic fields appearing in (\ref{supmagn}) in order to work
with elementary fields with mass dimension one. We then have a
$\mc{N}=1$ supersymmetric $SU(\tilde N)$ gauge theory with $N_f$ magnetic flavours $(q, \tilde q)$,
an adjoint field $Y$, and two gauge singlet mesons $M_1$,$M_2$, with canonical Kahler
potential. The superpotential, with rescaled couplings, reads
\bea
W_{magn}&=&\frac{g_Y}{3} \Tr Y^3+\frac{m_Y}{2} \Tr Y^2+\lambda_Y \Tr Y
+ \tr ( h_1 M_1 q \tilde q+h_2 M_2 q \tilde q + h_3 M_1 q Y \tilde q )  \non \\
\label{supmagnbis}
&&-h_1 m_{1}^2 ~ \tr M_1-h_2 m_{2}^2 ~ \tr M_2 + m_3 ~ \tr M_1^2
\eea
where the rescaled couplings in (\ref{supmagnbis}) are mapped to the original ones in (\ref{supmagn}) via
\be
h_1=-\frac{\tilde m_Y}{2 \mu^2} \left( \alpha_1 \Lambda \right) \gamma^2 \qquad
h_2=-\frac{\tilde g_Y}{\mu^2} \left( \alpha_2 \Lambda^2 \right) \gamma^2 \qquad
h_3= -\frac{\tilde g_Y}{\mu^2} \left( \alpha_1 \Lambda \right) \gamma^2 \beta  \non
\ee
\be
%
%
\label{rescale}
h_1 m_{1}^2=-m_Q \, \alpha_1 \, \Lambda \qquad 
h_2 m_{2}^2=-\lambda_Q \, \alpha_2\, \Lambda^2  \qquad 
m_3=h (\alpha_1 \Lambda)^2 
\ee
We can choose the magnetic quarks $q, \tilde q^{T}$ (which are $N_f \times \tilde N$ matrices)  
to solve the $D$ equations as
\be
\label{flavours}
q=\left ( \begin{array}{c}
k\\
0\\
\end{array}\right) \qquad \tilde q^{T} =\left ( \begin{array}{c}
\tilde k\\
0\\
\end{array}\right) 
\ee
where $k, \tilde k$ are $\tilde N \times \tilde N$ diagonal matrices
such that the diagonal entries satisfy $|k_i|=|\tilde k_i|$.

We impose the $F$ equations of motion for the superpotential (\ref{supmagnbis})
\bea
\label{Feqs}
F_{\lambda_Y}&=&\Tr Y=0 \non\\
F_Y&=&g_Y Y^2+m_Y Y+\lambda_Y +h_3 M_1 q\tilde q =0    \non   \\
F_q&=& h_2 M_2 \tilde q+h_1 M_1 \tilde q+h_3 M_1 Y \tilde q =0\non \\
F_{\tilde q}&=& h_2 M_2 q+h_1 M_1 q+h_3 M_1 q  Y =0\\
F_{M_1}&=& h_1 q \tilde q+ h_3 q Y \tilde q-h_1 m_{1}^2 \delta_{ij} +2 m_3 M_1=0 \qquad ~~i,j=1,\dots N_f    \non  \\ 
\label{rank}
F_{M_2}&=& h_2 q \tilde q-h_2 m_{2}^2 \delta_{ij}=0  \qquad \, \qquad ~~~~~~~~~~~~~~~~~~~~\,   i,j=1,\dots N_f 
\eea
Since we are in the range (\ref{range})
where $N_f>\tilde N$ 
the equation (\ref{rank}) 
is the rank condition of \cite{Seiberg}:
supersymmetry is spontaneously broken at tree-level by these non trivial 
$F$-terms. 

We can solve the first 
$\tilde N$ equations of (\ref{rank}) by
fixing the product $k \tilde k$ 
to be
$k \tilde k=m_2^2 \mathbf{1}_{\tilde N}$. 
We then parametrize
the quarks vevs in the vacuum (\ref{flavours}) with complex $\theta$
\be
\label{flavours2}
q=\left ( \begin{array}{c}
m_2 e^{\theta}\, \mathbf{1}_{\tilde N}\\
0\\
\end{array}\right) \qquad \tilde q^{T} =\left ( \begin{array}{c}
m_2 e^{-\theta}\, \mathbf{1}_{\tilde N}\\
0\\
\end{array}\right)~.
\ee
The other $N_f-\tilde N$ equations of (\ref{rank})
cannot be solved and so the corresponding $F$-terms don't
vanish ($F_{M_2}\neq 0$).
However we can find a vacuum configuration which satisfies all
the other $F$-equations (\ref{Feqs}) and the $D$-ones.
We solve the equations (\ref{Feqs}) for $M_1$, $Y$ and $\lambda_Y$ and we choose 
$Y$ to be diagonal,
finding
\be
\label{lambda}
\lambda_Y=\frac{h_3 h_1 m_2^2}{2 m_3} \left(m_2^2-m_1^2 \right)
-\frac{m_Y^2}{g_Y} \left( 1-\frac{h_3^2 m_2^4}{2 m_3 m_Y} \right)^2 \frac{n_1 n_2}{(n_1-n_2)^2}
\ee
where the integers $(n_1,n_2)$ count the eigenvalues degeneracy
along the $Y$ diagonal,
with $(n_1+n_2=\tilde N)$
\be
\non
\langle Y \rangle= 
\left( \begin{array}{c  c}
y_1 \mathbf{1}_{n_1}&0\\
0&y_2\mathbf{1}_{n_2} \\
\end{array} \right)
\qquad 
y_1=-\frac{m_Y-\frac{h_3^2 m_2^4}{2 m_3}}{g_Y} \frac{n_2}{n_1-n_2} 
\qquad
y_2=\frac{m_Y-\frac{h_3^2 m_2^4}{2 m_3}}{g_Y} \frac{n_1}{n_1-n_2}
\ee
We choose the vacuum in which
the magnetic gauge group is not broken by
the adjoint field choosing $n_1=0$, so $y_2$ vanishes
and $\langle Y \rangle=0$. We observe
that other choices for $\langle Y \rangle$ with $n_1 \neq 0 \neq n_2$
wouldn't change the tree-level potential energy of the vacua
which is given only
by the non vanishing $F_{M_2}$. 
This classical landscape of
equivalent vacua will be wiped out by 1-loop quantum corrections\footnote{
This agrees with an observation in \cite{ooguri}.}.
In our case ($n_1=0$) we have 
\be
\label{vuotometa}
\langle M_1 \rangle=\left( \begin{array}{c  c}
\frac{h_1}{2 m_3} (m_1^2-m_2^2 )~ \mathbf{1}_{\tilde N}&0\\
0&\frac{h_1 m_1^2}{2 m_3}~\mathbf{1}_{N_f-\tilde N}\\
\end{array} \right)=
\left( \begin{array}{c  c}
p_1^A&0\\
0&p_1^B\\
\end{array} \right)
\ee 
The two non trivial blocks are respectively $\tilde N$ and $N_f-\tilde N$
diagonal squared matrices.

The $(q,\tilde q)$ $F$ equations fix the vev of the $M_2$ meson
to be
\be
\label{vuotometa2}
\langle M_2 \rangle=\left( \begin{array}{c  c}
-\frac{h_1^2}{2 h_2 m_3} (m_1^2-m_2^2 )~ \mathbf{1}_{\tilde N}&0\\
0&\mc{X}\\
\end{array} \right)
=\left( \begin{array}{c  c}
p_2^A&0\\
0&\mc{X}\\
\end{array} \right)
\ee 
where the blocks have the same dimensions of $M_1$,
with $\mc{X}$ undetermined at the classical level. 

Since supersymmetry is broken at tree level by the rank condition (\ref{rank})
the minimum of
the scalar potential is
\be
\label{Vmin}
V_{MIN}=|F_{M_2}|^2=(N_f-\tilde N)|h_2 m_2^2|^2=(N_f-\tilde N)\,  \alpha_2^2 \, |\lambda_Q \Lambda^2|^2
\ee
It depends on parameters that we can't compute from the electric theory
(e.g. $\alpha_2$); 
in any case we are only interested in the
qualitative behaviour of the non supersymmetric state.
The potential energy of the vacuum (\ref{Vmin}) 
doesn't depend on 
$\theta$ and $\mc{X}$;
they are massless fields at tree level,
not protected by any symmetry 
and hence are pseudomoduli.
Their fate will be decided by the quantum corrections.

We don't expect the value of $\mc{X}$ in the quantum minimum to
vanish because there isn't any $U(1)_R$ symmetry.
Indeed, computing the 1-loop corrections, we will find that in the 
quantum minimum 
the value of $\theta$ is zero
while $\mc{X}$ will get a nonzero vev. 
This makes our metastable minimum different from the
one discovered in \cite{Seiberg,Uranga,ooguri} where the quantum corrections
didn't give the pseudomoduli a nonzero vev.
Notice also that although we have many vevs different from zero in the
non supersymmetric vacuum they are all smaller than the natural breaking mass scale
$|F_{M_2}|^{\frac{1}{2}}=|h_2 m_2^2|^{\frac{1}{2}}$.

\section{1-Loop effective potential}
In this section we study the 1-loop quantum corrections to the
effective potential for the fluctuations around the 
non supersymmetric vacuum
selected in the previous section with $\langle Y \rangle=0$.
The aim is to estabilish the sign of the mass corrections for
the pseudomoduli $\mc{X},\theta$. 
The 1-loop corrections to the tree level potential energy 
depend on the choice of the adjoint vev 
$\langle Y \rangle$: as a matter of fact they are minimized
by the choice $\langle Y \rangle=0$.

The 1-loop contributions of
the vector multiplet to the effective potential vanish
since the $D$ equations are satisfied by 
our non supersymmetric vacuum configuration.

The 1-loop corrections will be computed using
the supertrace of the bosonic and fermionic 
squared mass matrices built up from the
superpotential for the fluctuations of the fields
 around the vacuum.
The standard expression of the 1-loop effective potential
is
\be
\label{supertrace}
V_{1-loop}=\frac{1}{64 \pi^2} S\Tr \mc{M}^4 \log \frac{ \mc{M}^2}{\Lambda^2}=
\frac{1}{64 \pi^2} \sum \left( m_B^4 \log \frac{ m_B^2}{\Lambda^2} - m_F^4 \log \frac{ m_F^2}{\Lambda^2} \right)
\ee
where the $F$ contributions to the mass matrices are read from the superpotential $W$
\be
\label{massmat}
m^2_B=\left( \begin{array}{c c}
W^{\dagger a c}W_{cb}&W^{\dagger a b c}W_{c}\\
W_{a b c}W^{\dagger c}&W_{a c}W^{\dagger cb}\\
\end{array}
\right)
\qquad
m^2_f= 
\left(
\begin{array}{c c}
W^{\dagger a c}W_{cb}&
0 \\
0&
W_{a c}W^{\dagger cb} \\
\end{array}
\right)
\ee 
We parametrize the fluctuations around the tree level vacuum as
\be
\label{para1}
q=\left ( \begin{array}{c}
k e^{\theta}+\xi_1\\
\phi_1\\
\end{array}\right) 
\qquad 
\tilde q^{T} =\left ( 
\begin{array}{c}
k e^{-\theta}+ \xi_2\\
\phi_2\\
\end{array}\right)
\qquad
Y=\delta Y 
\ee
\be
\label{para2}
M_1=\left ( \begin{array}{c c}
p_1^A+\xi_3&\phi_3\\
\phi_4&p_1^B+\xi_4\\
\end{array}\right) 
\qquad
M_2=\left ( \begin{array}{c c}
p_2^A+\xi_5&\phi_5\\
\phi_6&\mc{X}\\
\end{array}\right)
\ee 
We expand the classical superpotential (\ref{supmagnbis})
up to trilinear order in the fluctuations $\phi_i,\xi_i,\delta Y$.
Most of these fields acquire tree level masses, but there are
also massless fields. 
Some of them are goldstone bosons of the global symmetries 
considering $SU(\tilde N)$ 
global, the others are pseudogoldstone.\\ 
In this set up, 
$\xi_1$ and $\xi_2$ combine to give the same goldstone 
and pseudogoldstone bosons as in \cite{Seiberg}. Gauging the $SU(\tilde N)$
symmetry these goldstones are eaten by the vector fields,
and the other massless fields, except $\theta+\theta^{\star}$, 
acquire positive masses from $D$-term potential as in \cite{Seiberg}.
Combinations of the $\f_i$ fields give the
goldstone bosons related to the breaking of the flavour symmetry
$SU(N_f)\to SU(\tilde N)\times SU(N_f-\tilde N)\times U(1)$.
The off diagonal elements of the classically massless field $\mc{X}$
are goldstone bosons of the $SU(N_f-\tilde N)$ flavour symmetry as in \cite{Uranga}. 
We then end up with the pseudogoldstones $\theta+\theta^{\star}$
and the diagonal $\mc{X}$. 

We now look for the fluctuations which give  
contributions to the mass matrices (\ref{massmat}). They are only the
$\f_i$ fields, while the $\xi_i$ and $\delta Y$ represent
a decoupled supersymmetric sector. Indeed $\xi_i$ and $\delta Y$ 
do not appear in bilinear terms coupled to the $\f_i$ sector,
so they do not contribute to the fermionic mass matrix (\ref{massmat}). 
Even if they appear in trilinear terms coupled to the $\f_i$, they
do not have the corresponding linear term\footnote{The possible linear terms in 
$\xi_i$ and $\delta Y$ factorize the $F$-equations (\ref{Feqs}) and so they all vanish.}:
they do not contribute to the bosonic mass matrix (\ref{massmat}).
Since $(\xi_1,\xi_2,\delta Y)$ do not couple to
the breaking sector at this order, also
their $D$-term contributions to the mass matrices vanish and all of them 
can be neglected. 
We can then restrict ourselves to the 
chiral $\f_i$ fields
for computing the 1-loop quantum corrections to the effective 
scalar potential using (\ref{massmat}).
Without loss of generality we can set the pseudomoduli $\mc{X}$
proportional to the identity matrix.

The resulting
superpotential for the sector affected by the supersymmetry
breaking (the $\phi_i$ fields) 
is a sum of $\tilde N \times (N_f -\tilde N)$ decoupled copies of a
model of chiral fields
which breaks supersymmetry at tree-level
\bea
W&=&h_2 \left( \mathcal{X} \f_1 \f_2-m_2^2 \mc{X} \right)+h_2 m_2 \left( e^{\theta} \f_2 \f_5
+ e^{-\theta} \f_1 \f_6 \right) + \non \\
\label{orafe2} 
&&+ h_1 m_2 \left( e^{\theta} \f_2 \f_3+ e^{-\theta} \f_1 \f_4 \right) + 2 m_3 \f_3 \f_4 + \frac{h_1^2 m_1^2}{2 m_3} \,  \f_1 \f_2
\eea
This superpotential doesn't have any $U(1)_R$ symmetry, differently from
the ones studied in
\cite{Seiberg,Uranga,ooguri}. This may be read as an example of 
a non generic superpotential which
breaks supersymmetry \cite{Nelson}, without exact $R$ symmetry. 
 
The expressions for the eigenvalues, and then for the 
1-loop scalar potential, are too complicated to be written here.
We can plot our results numerically to give
a pictorial rapresentation.

The computation is carried out in this way: we first compute the eigenvalues of the
bosonic and fermionic mass matrices (\ref{massmat}) using
the superpotential (\ref{orafe2}); we evaluate them
where all the fluctuations $\f_i$ are set to zero;
finally we compute the 1-loop scalar potential using (\ref{supertrace})
 as a function of the pseudomoduli $\mc{X},\theta+\theta^{\star}$. 
The corrections will always be powers of
$\theta+\theta^{\star} \equiv \tilde \theta$ so from now on we will treat only the
$\tilde \theta$ dependence.
We give graphical plots of the 1-loop effective potential treating fields and
couplings as real.
We have checked that our qualitative conclusions
about the stability of the vacuum are not affected by
using complex variables.

We redefine the couplings in order to have the mass matrices
as functions of three dimensionless parameters ($\rho,\eta,\zeta$)
\be
\label{parameters}
\rho=\frac{h_1}{h_2} \qquad \eta=\frac{2 m_3}{h_2 m_2} \qquad \zeta= \frac{h_1^2 m_1^2}{2 h_2 m_2 m_3}~~,
\qquad \qquad   \zeta<\rho<\eta
\ee
and we rescale the superpotential with an overall scale
$h_2 m_2$ which becomes the fundamental unit of our
plots.
The inequality in (\ref{parameters}) is a consequence
of the range (\ref{rangemasses}) and the redefinitions (\ref{rescale}).
We notice also that $(\rho,\eta,\zeta)$ have absolute
values smaller than one.

\begin{figure}[htbc]
\begin{center}
\begin{tabular}{cc}
\includegraphics[width=8cm, height=7cm]{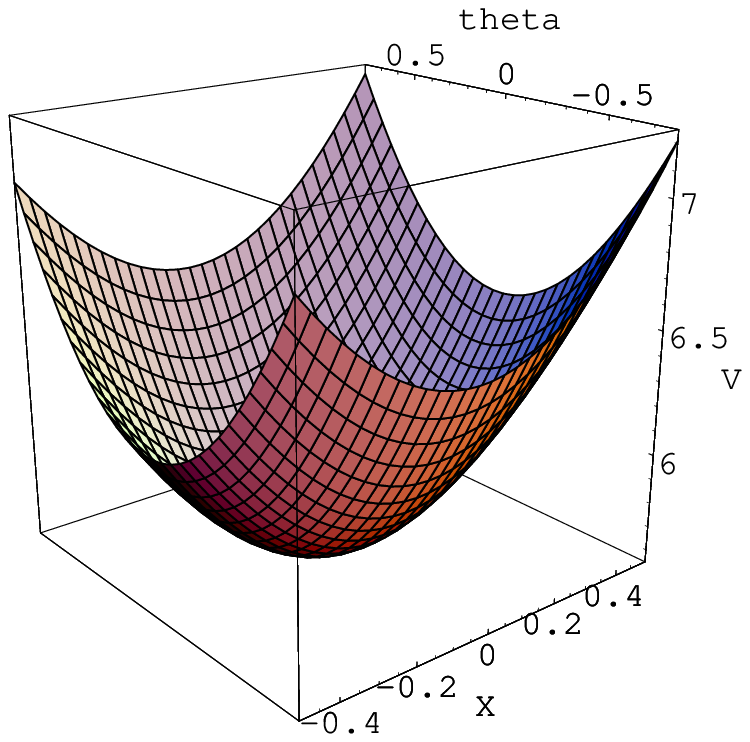}
\includegraphics[width=8cm, height=5cm]{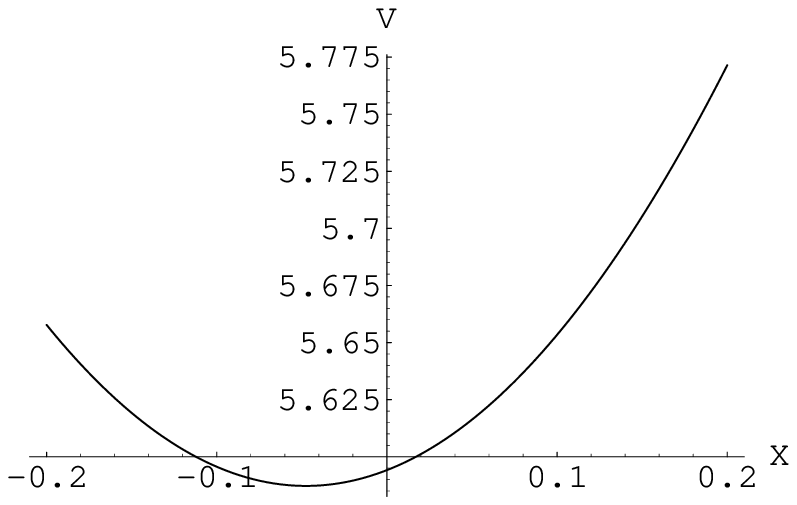}
\end{tabular}
\caption{Scalar potential $V^{1-loop}$ for ($\eta=0.5, \rho=0.1, \zeta=0.05, 
\mc{X}=-0.5 \dots 0.5, 
\tilde \theta=-0.8\dots 0.8$), and its section for $\tilde \theta=0$;
$\mc{X}$ is in unit of $m_2$, while $V$ is in unit of $|h_2^2 m_2^2|^2$.}
\label{vtetax}
\end{center}
\end{figure}

In figure \ref{vtetax} we plot the 1-loop scalar potential as a function of
$\mc{X},\tilde \theta$ and for
fixed values of the parameters $\rho,\eta,\zeta$.
We can see that there is a minimum, so the moduli space is lifted 
by the quantum corrections, the pseudomoduli get positive masses, and there
is a stable non supersymmetric vacuum. 
Making a careful analysis we find that
the quantum minimum in the 1-loop scalar potential
is reached when $\langle \tilde \theta \rangle=0$ but
$\langle \mc{X} \rangle \neq 0$ and its
vev in the minimum depends on the parameters ($\rho,\eta,\zeta$).
This agrees with what we observed in the previous section.
It can be better seen in the second picture of figure \ref{vtetax}
where we take a section of the first plot for $\tilde \theta=0$.

\begin{figure}[htbc]
\begin{center}
\includegraphics[width=8cm, height=6cm]{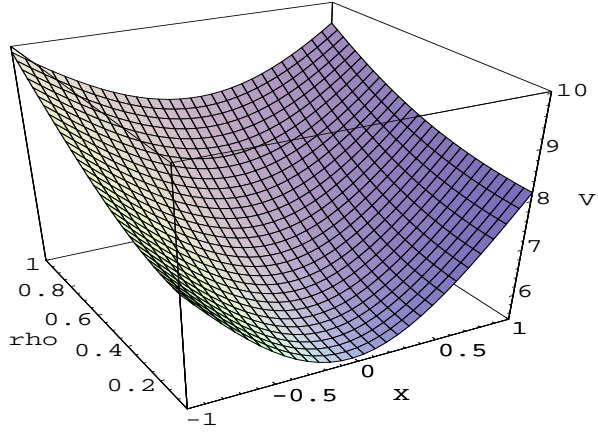}
\caption{Scalar potential $V^{1-loop}$ for ($\mc{X}=-1\dots 1,\rho=0.05 \dots 1, \eta=0.5,\zeta=0.05,\tilde \theta=0$);
$\mc{X}$ is in unit of $m_2$, while $V$ is in unit of $|h_2^2 m_2^2|^2$.}
\label{v1loopux}
\end{center}
\end{figure}

In figure \ref{v1loopux} we plot the 1-loop
scalar potential for $\tilde \theta=0$ as a function of
$\mc{X}$ and of the parameter $\rho$, fixing $\eta$ and $\zeta$.
For each value of $\rho$ the curvature around the
minimum gives a qualitative estimation of the
generated mass for the pseudomoduli $\mc{X}$.
We note that for large $\rho$ the scalar potential
become asymptotically flat, and so the 1-loop generated mass
goes to zero, but this is outside our allowed range.

As already observed, there is a minimum 
for $\langle \mc{X} \rangle$ slightly different from zero
due to quantum corrections, and we have found that it goes to
zero in the limit ($\zeta \to 0,\rho \to 0$).
We can give analytic results in this limit\footnote{Considering $\eta,\rho,\zeta$ real.}.
We found at zero order in $\rho$ and $\zeta$, with arbitrary $\eta$,
that the 1-loop generated masses
for the pseudomoduli are
\bea
\label{masses1}
m_{\mc{X}}^2 &=& \frac{\tilde N (N_f-\tilde N)}{8\p^2}|h_2^2 m_2|^2 (\log [4]-1)+ o(\rho)+o(\zeta) \\
m_{\tilde \theta}^2 &=&\frac{\tilde N (N_f-\tilde N)}{16 \p^2}|h_2^2 m_2^2|^2 ( \log [4]-1)+o(\rho)+o(\zeta) \non
\eea
so in the limit of small $\rho$ (and small $\zeta$) quantum corrections
don't depend on $\eta$.
We can write the 1-loop scalar potential
 setting $\eta$ to zero
obtaining (for small $\zeta$)
\bea
\label{V1loop}
V^{(1)}&=&\frac{\tilde N (N_f-\tilde N)}{64 \p^2} |h_2^2 m_2|^2 \bigg \{ |m_2|^2 \bigg(\log \big( \frac{|h_2 m_2|^2}{\Lambda^2}
\big)
+2 \rho^4 \log[\rho^2]-4 (1+\rho^2)^2 \log[1+\rho^2]+ \non\\ 
&&+ 2 (2+\rho^2)^2 \log[2+\rho^2]\bigg)+ \bigg(4 (2+\rho^2)^2 \log[2+\rho^2]-4 \rho^4 \log[\rho^2]+ \\
&&-8 (1+\rho^2)(1+2 \log[1+\rho^2])\bigg) |\mc{X}+m_2 \zeta|^2+ |m_2|^2 \bigg(2(1+\rho^2)\bigg[ (2+\rho^2)^2 \log[2+\rho^2]+ \non\\
&&-\rho^4 \log[\rho^2]-2(1+\rho^2)(1+2 \log[1+\rho^2])\bigg] 
+4( \log[4]-\frac{5}{3})\zeta^2 \bigg)(\theta+\theta^{\star})^2  \bigg\} (1+o(\zeta))\non
\eea
where this expression is valid only in the regime of small $\rho$. 
In these approximations the vev for $\langle \mc{X}\rangle$ in the minimum
is shifted linearly with $\zeta$; however, in general, the complete behaviour 
for $\langle \mc{X}\rangle$ is more complicated and depends non trivially
on $\eta$. We observe that, being $\zeta$ a simple shift for the vev of $\mc{X}$, it doesn't affect
its mass, while it modifies $\tilde \theta$ mass.

From (\ref{V1loop}) we can
read directly the masses expanding for small $\rho$
\bea
\label{massX}
m_{\mc{X}}^2&=&\frac{\tilde N (N_f -\tilde N)}{8 \p^2}|h_2 m_2|^2 \bigg(|h_2|^2 (\log[4]-1)+|h_1|^2 (\log[4]-2)
\bigg)
\\
\label{masstheta}
m_{\tilde \theta}^2&=&\frac{\tilde N (N_f -\tilde N)}{16 \p^2}|h_2 m_2^2|^2 \bigg(|h_2|^2 (\log[4]-1)+\left| 
\frac{h_1^2 m_1^2}{2 m_2 m_3}\right|^2 (\log[4]-\frac{5}{3})+\non \\
&&~~~~~~~~~~~~~~~~~~~~~~~~~~~~~~~~~~~+|h_1|^2(2\log[4]-3)
\bigg) \, .
\eea
These expressions are valid up to cubic order in $\rho,\zeta$.
The first term in (\ref{massX},\ref{masstheta}), being independent
of the deformations ($\rho,\eta,\zeta$), agrees with \cite{Seiberg}.
The second term in (\ref{massX}) is the same as in 
\cite{ooguri}.

\section{Supersymmetric vacuum}
Supersymmetry is restored via non perturbative effects \cite{Affleck}, 
away from the metastable vacuum
in the field space,
when the $SU(\tilde N)$ symmetry is gauged
\cite{Seiberg}.
The non supersymmetric vacuum discovered in the
sections 3 and 4 is a metastable state of the theory
which decays to a supersymmetric one.
We are interested in evaluating the lifetime of the metastable
vacuum. 
We need an estimation of the vevs of the elementary 
magnetic fields
in the supersymmetric state.

We first integrate out the massive fields in 
the superpotential (\ref{supmagnbis}) using their equations of motion.
In (\ref{supmagnbis}) there are two massive fields ($M_1,Y$). 
We integrate out the meson $M_1$ and the adjoint field $Y$ tuning $\lambda_Y$ in such a way that
the gauge group $SU(\tilde N)$ is not broken by the adjoint\footnote{
We are not interested in finding all the supersymmetric vacua.}, as in the metastable
state, so $\langle Y \rangle=0$.
Using this last condition
the equation of motion for the meson $M_1$ gives the simple relation 
$M_1=\frac{h_1}{2 m_3} \left(m_1^2- q \tilde q \right)$.
Integrating out the charged field $Y$ the scale matching condition reads
\be
\tilde \Lambda^{2 \tilde N-N_f}=\tilde \Lambda_{int}^{3 \tilde N-N_f} \hat m_Y^{-\tilde N}
\ee
where we have indicated with $\hat m_Y$ the resulting mass for $Y$ which
is a combination of its tree-level mass $m_Y$ and a term proportional to
$\frac{h_3^2}{m_3} (q \tilde q)^2$ which will be shown to be
zero in the supersymmetric vacuum.

We obtain a superpotential for the meson $M_2$ and the 
flavours $(q, \tilde q)$
\be
\label{sup12}
W_{int}=\tr \left( \frac{h_1^2}{4 m_3} \left(2 m_1^2 q \tilde q- (q\tilde q)^2 \right)+ h_2 M_2 q \tilde q
-h_2 m_2^2 M_2   \right)
\ee
We expect that the supersymmetric vacua lie in the
large field region, where the $SU(\tilde N)$ gauge dynamics 
becomes relevant \cite{Seiberg}.
We then consider large expectation value for the meson $M_2$.
We can take as mass term for the flavours $(q, \tilde q)$
only the vev $h_2 \langle M_2 \rangle$ neglecting the other contribution in (\ref{sup12})
coming from the couplings
of the magnetic theory.

We then integrate out the flavours $(q, \tilde q)$ using their equations of motion
$(q=0,\tilde q=0)$. The corresponding scale matching condition is
\be
\label{scale1}
\Lambda_L^{3 \tilde N}=\tilde \Lambda_{int}^{3 \tilde N-N_f} \det \left(h_2 M_2\right)=
\tilde \Lambda^{2 \tilde N-N_f} \det \left(h_2 M_2\right) m_{Y}^{\tilde N} \, .
\ee

The low energy effective $SU(\tilde N)$ superpotential gets a non-perturbative contribution from
the gauge dynamics related to the gaugino condensation proportional
to the low energy scale $\Lambda_L$
\be
W=\tilde N \Lambda_{L}^{3}
\ee
that can be written in terms of the macroscopical 
scale $\tilde \Lambda$ using (\ref{scale1}). This contribution should be
added to the $M_2$ linear term that survives in (\ref{sup12}) after having integrated out
the magnetic flavours ($q, \tilde q$). 
Via the scale matching relation (\ref{scale1}) we can then express
the low energy effective superpotential as a function of only the $M_2$ meson 
\be
\label{Wlow}
W_{Low}=\tilde N \left(\tilde \Lambda^{2 \tilde N-N_f} \det (h_2 M_2) \right)^{\frac{1}{\tilde N}} m_Y - m_2^2 h_2 \, \tr M_2
\ee

Using this dynamically generated superpotential we can obtain the 
vev of the meson $M_2$ in the supersymmetric vacuum.
Considering $M_2$ proportional to the identity $\mathbf{1}_{N_F}$ we minimize (\ref{Wlow})
and obtain 
\be
\label{M_2}
h_2 \langle M_2 \rangle=\tilde \Lambda \epsilon^{\frac{\tilde N}{N_f-\tilde N}} \xi^{\frac{\tilde N}{N_f-\tilde N}}\mathbf{1}_{N_f}
=m_2 \left( \frac{1}{\epsilon}\right)^{\frac{N_f-2 \tilde N}{N_f-\tilde N}} \xi^{\frac{\tilde N}{N_f-\tilde N}} \mathbf{1}_{N_f}
\ee
where
\be
\epsilon=\frac{m_2}{\tilde \Lambda} \qquad \xi=\frac{m_2}{m_Y} \,.
\ee
$\epsilon$ is a dimensionless parameter which can be made parametrically
small sending the Landau pole $\tilde \Lambda$ to infinity.
$\xi$ is a dimensionless finite parameter which 
doesn't spoil our estimation of the supersymmetric vacuum in the sensible
range $\epsilon<\frac{1}{\xi}$.
All the exponents appearing in (\ref{M_2}) are positive in our window (\ref{range}). 

We observe that in the small $\epsilon$ limit the vev $h_2 \langle M_2 \rangle$ is larger than the
typically mass scale $m_2$ of the magnetic theory 
but much smaller than
the scale $\tilde \Lambda$
\be
m_2 \ll h_2 \langle M_2 \rangle \ll \tilde \Lambda.
\ee
This fact justifies our approximation in integrating out the massive flavours $(q, \tilde q)$
neglecting the mass term in (\ref{sup12}) except $h_2 \langle M_2 \rangle$. It also
shows that the evaluation of the supersymmetric vacuum is reliable because the 
scale of $h_2 \langle M_2 \rangle$ is well below the Landau pole.

\section{Lifetime of the metastable vacuum}
We make a qualitative evaluation of the decay rate of the metastable vacuum. 
At semi classical level the decay probability is proportional to
$e^{-S_B}$ where $S_B$ is the
bounce action from the non supersymmetric vacuum to a supersymmetric one.
We have to find a trajectory in the field space such that the potential 
energy barrier is minimized.
We remind the non supersymmetric vacuum configuration (\ref{flavours2},\ref{vuotometa},\ref{vuotometa2})
and the supersymmetric one 
\be
\label{susyvac}
q=0 \qquad \tilde q=0 \qquad Y=0 \qquad \langle M_1 \rangle=\frac{h_1 m_1^2}{2 m_3} ~~ \mathbf{1}_{N_f}  \qquad \langle M_2 \rangle \neq 0
\ee
where $\langle M_2 \rangle$ can be read from (\ref{M_2}).

By inspection of the $F$-term contributions (\ref{Feqs}) to the
potential energy it turns out
that
the most efficient path is to climb from the local non supersymmetric minimum to the
local maximum where all the fields are set to zero but for $M_1$ which
has the value $M_1= \frac{h_1 m_1^2}{2 m_3} \mathbf{1}_{N_f}$ as 
in the supersymmetric vacuum,
and $M_2$, which is as in (\ref{vuotometa2}).
This local maximum has potential energy
\be
V_{MAX}=N_f |h_2 m_2^2|^2
\ee
We can move from the local maximum to the supersymmetric minimum (\ref{susyvac}) 
along the $M_2$ meson direction. 
The two 
minima are not of the same order and so the thin wall 
approximation of \cite{Coleman} 
can't be used. 
We can approximate
the potential barrier with a triangular one
using the formula of \cite{Duncan}
\be
S\simeq\frac{(\Delta \Phi)^4}{V_{MAX}-V_{MIN}}
\ee
We neglect the difference in the field space
between all the vevs at the non supersymmetric vacuum and
at the local maximum.
We take as $\Delta \Phi$ the difference between the
vevs of $M_2$ at the local maximum and at the
supersymmetric vacuum.
Disregarding the $M_2$ vev at the local maximum we can
approximate $\Delta \Phi$ as (\ref{M_2}).
We then obtain as the decay rate
\be
S\sim \left(\left( \frac{1}{\epsilon}\right)^{\frac{N_f-2 \tilde N}{N_f-\tilde N}} \xi^{\frac{\tilde N}{N_f-\tilde N}} \right)^4
\sim \left( \frac{1}{\epsilon}\right)^{4 \frac{N_f-2 \tilde N}{N_f-\tilde N}}
\ee
This rate 
can be made parametrically large sending to zero the dimensionless
ratio $\epsilon$ (i.e. sending $\tilde \Lambda \to \infty$)
 since the exponent $ \left( 4 \frac{2 \tilde N-N_f}{\tilde N- N_f}\right)$ is always
positive in our window (\ref{range}).\\

In conclusion we have found that the $SU(N_c)$ SQCD with two adjoint chiral
fields and mesonic deformations admits a metastable
non supersymmetric vacuum with parametrically long life.
It seems that particular care is needed in building
models with adjoint matter exhibiting such
vacua. The same can be said about the
string geometrical construction 
realizing the gauge model we have studied \cite{Giveon,Giveon1}.

\section{Acknowledgments}
We would like to thank A.Zaffaroni and F.Zwirner
for discussions. \\
This work has been supported in part by INFN, PRIN prot.2005024045-002 and the
European Commission RTN program MRTN-CT-2004-005104.

\end{document}